\documentclass[sigconf, screen]{acmart}

\AtBeginDocument{%
  }

\setcopyright{none}
\acmYear{2024}
\acmDOI{}

\acmConference[FAccTRec'24]{7th FAccTRec Workshop on Responsible Recommendation at RecSys 2024, Bari, Italy}{October 14, 2024}{Bari, Italy}
\acmISBN{}

\begin{document}

\title{Questioning AI: Promoting Decision-Making Autonomy Through Reflection}

\author{Simon W.S. Fischer}
\email{simon.fischer@donders.ru.nl}
\orcid{0000-0003-2992-6563}
\affiliation{%
  \institution{Donders Institute for Brain, Cognition, and Behaviour}
  \city{Nijmegen}
  \country{Netherlands}}


\received{16 September 2024}

\maketitle

\section{Introduction}
Decision-making is increasingly supported by machine recommendations. In healthcare, for example, a clinical decision support system is used by the physician to find a treatment option for a patient. In doing so, people can rely too much on these systems, which impairs their own reasoning process. 

The European AI Act addresses the risk of over-reliance and postulates in Article 14 on human oversight that people should be able ``to remain aware of the possible tendency of automatically relying or over-relying on the output'' \cite{AIAct}. Similarly, the EU High-Level Expert Group identifies human agency and oversight as the first of seven key requirements for trustworthy AI \cite{HLEG2019}. 

The following position paper proposes a conceptual approach to generate machine questions about the decision at hand, in order to promote decision-making autonomy. This engagement in turn allows for oversight of recommender systems.
The systematic and interdisciplinary investigation (e.g., machine learning, user experience design, psychology, philosophy of technology) of human-machine interaction in relation to decision-making provides insights to questions like: \textit{how to increase human oversight and calibrate over- and under-reliance on machine recommendations; how to increase decision-making autonomy and remain aware of other possibilities beyond automated suggestions that repeat the status-quo}?

\section{Shortcomings of Explanations}

To ensure oversight, Article 13 of the European AI Act requires the ability to ``interpret a system’s output and use it appropriately'' \cite{AIAct}. The assumption is that the increased transparency of the system's behaviour allows operators (i.e., decision-makers) to better understand and monitor the system, which ideally contributes to a more informed decision-making. One solution to increase transparency of recommender systems is explainable AI (XAI). Explanations are important to sense-making, but it is not trivial to provide them, as different people require different explanations depending on the context \cite{Zednik2019}. As such, physicians require different explanations than patients, more so than model developers. The agent-relative nature of explanations leads to at least two interrelated problems.

First, there is a gap between XAI methods and context-specific needs \cite{Liao2020}. Current XAI methods, with their algorithm-centred view, are primarily aimed at model developers and researchers, but not at operators or decision-makers, who have less technical knowledge and need situational information \cite{Miller2019,Ehsan2021}.
Second, explanations do not help to sufficiently calibrate reliance on machines. Explanations are either ignored (i.e., under-reliance) \cite{VanDerWaa2021,Burton2020}, or they can reinforce a decision regardless of its quality (i.e., over-reliance) \cite{Bansal2021,Jacobs2021}. The mere provision of more information in favour of transparency, which, however, does not meet the epistemic needs of the decision-maker, therefore does not automatically lead to the appropriate use and oversight of recommender systems.

\subsection{Related Work}

There is a small but growing body of literature that discusses various ways to engage the decision-maker in the decision-making process that involves machine recommendations and explanations. The overarching aim is to calibrate the reliance on these systems (i.e., avoid over- and under-reliance) \cite{Naiseh2021}. In one study, for example, different ways of presenting explanations, i.e.,  cognitive forcing strategies, were investigated: on demand; delayed after 30 seconds; and only once after the human had made their decision. The results show that these interventions reduce over-reliance on machine explanations, but the most effective were also the least favoured by participants \cite{Bucinca2021}. Another paper proposes a hypothesis-driven recommender system that shows evidence for and against skin cancer, so that the physician can make an informed decision rather than simply accepting or rejecting a recommendation
\cite{Miller2023}. Although a different study showed that presenting evidence for and against the diagnosis did not significantly improve diagnostic accuracy, the reflective aspect of this approach was appreciated by physicians \cite{Cabitza2023}. 

\section{Engagement through Questions}
To further develop this line of research to increase engagement and thus human oversight over machine recommendations, this paper proposes the concept of a question-asking machine \cite{Haselager2023}. This additional system (or component of an existing recommender system) asks the decision-maker questions to encourage critical thought and reflection on the decision at hand. Although this system has not yet been fully implemented and evaluated, another study found that framing causal explanations as questions can improve human judgement \cite{Danry2023}.

To formalise evidence and hypotheses and generate meaningful machine questions, we need to combine case data, machine recommendations, XAI methods, and human domain knowledge in a probabilistic approach. The aim is to bridge the above-mentioned gap between explanations and context-specific needs to create more meaningful human-computer interaction with appropriate reliance. 

\subsection{Case Study}

An illustrative example is the medical field, where a doctor uses a clinical decision support system to find the best treatment option for already diagnosed patients.
A physician has a professional responsibility to promote the health and well-being of the patient. This includes an epistemological responsibility of the physician for how they make a diagnosis and what treatment they suggest \cite{vanBaalen2015}. In other words, the physician needs to be able to provide reasons for their decisions, instead of merely referring to the machine recommendation. Remaining critical towards the output is therefore crucial. 
 
Questions can be about the relevance and adequacy of the importance or significance of certain features that the recommender system considers. Next to using XAI methods like LIME or SHAP to provide feature importance, an additional step is added to help the physician reflect on and contextualise these probabilities.
Example questions could be ``Is symptom \textit{x} that contributed 62\% to the outcome actually that relevant?'' or ``How does diagnosis \textit{Y} follow from symptom \textit{x}?''. This type of question examines the apparent causal dependencies or associations of symptoms embedded in the model. As different risks or diseases can have common symptoms, and patient information comes from heterogeneous sources, it is relevant to consider different alternatives. Especially as an explanation for only the most likely option can be misleading \cite{Rudin2019}. As such, other questions can be about the strongest evidence for the alternative decision, and whether these symptoms can be easily ignored. 

Next, questions can be asked about scenarios in which the data looks different, similar to counterfactual explanations. To do so we need to perturb the input variables and simulate different scenarios. It can then be inquired whether a particular symptom could be reduced or increased by another smaller intervention to make the actual treatment more effective (e.g., ``Is it possible to change symptom \textit{x} from \textit{p} to \textit{q}? This would increase the chances of a positive treatment by 25\%''). The patient's expectation of recovery, for example, has a major influence on the treatment outcome. Besides, considering different input also allows the physician to reflect on and challenge built-int thresholds (e.g., ``If the patient was 3 years older, which reduces the chances of a positive result, would you still recommend the same treatment?'').

Other questions might relate to information that is not taken into account by the recommender system. This could be particularly important for diseases that are less well documented, or where the data distribution is skewed due to sampling bias, or simply because the information is not yet available because of an unperformed test (e.g., medical imaging).
\balance

Overall, the aim is to promote the physician's own judgement by critically questioning the machine recommendation and considering other hypotheses. In other words, instead of providing a recommendation (i.e., an answer), we want to make people aware of the range of possible answers in order to create solutions that fit the current case. The physician can benefit from data-based findings and at the same time re-contextualise case-related information. 
In this way, the physician can form a better picture of the patient, ensuring that the patient is treated in the most appropriate way, which has an impact on the patient's safety.
In addition, increasing the chances of a positive outcome is linked to the impact on fairness in relation to the patient's treatment. Many systems use erroneous or distorted data, resulting in underrepresented diseases being misdiagnosed and incorrectly treated. Questions can sensitise the physician to embedded assumptions and values in order to counteract potential tunnel vision. Moreover, questions can also be used to elicit the needs and values of the patient, e.g., specific preferences for certain treatment options or other personal circumstances. This could serve more as a fallback option, as the physician will ideally inquire about this information.  

Questions enable a more interactive interaction between the physician and the recommender system. Ideally, the recommender system could process the answer or feedback of the physician in order to adjust the output accordingly. For example, during the consultation, the physician realises that the patient has given a pain score in a screening questionnaire, which serves as input for the recommendation system, that is far above the appropriate value. Being able to adjust the input value and see how the output changes enhances the physician's understanding of the model's behaviour. More importantly, this interactivity increases the decision-making autonomy of the physician, thereby avoiding over-reliance on machine recommendations. 
The interactivity between recommender system and physician in turn improves the interaction between physician and patient, as the physician can communicate and discuss how they reached a decision.

\section{Concluding Remarks}

The proposed approach of posing questions about machine recommendations aims to put human expertise at the centre of decision-making while benefiting from data-driven insights from recommender systems. Particularly in domains such as medicine or law, where the decision-maker and decision-subject are different persons, epistemological responsibility, i.e., how a decision is made, is of crucial importance. 
The assumption of our approach is that asking questions increases engagement and avoids over-reliance on machine recommendations, which allows the decision-maker to give a more complete account of the decision-making process. 
It will be crucial to investigate what makes questions meaningful, how often and when to ask questions and when to simply provide information. These requirements highly depend on context. We thus also want to refrain from the implementation of large language models, as the relevance of the outputs bear the risk of being too unreliable. At a time when recommender systems are being increasingly deployed in decision-making processes, it is important to remain aware of other alternatives for which critical reflection on machine recommendations and explanations is necessary.

\begin{acks}
This research is supported by the Donders Institute for Brain, Cognition, and Behaviour.
\end{acks}

\bibliographystyle{ACM-Reference-Format}
\bibliography{sample-base}

\appendix

\end{document}